\documentclass[useAMS, usenatbib, onecolumn]{mn2e}
\usepackage{graphicx}
\usepackage{amssymb}

\newcommand\be{\begin{equation}}
\newcommand\ee{\end{equation}}
\newcommand\ba{\begin{eqnarray}}
\newcommand\ea{\end{eqnarray}}

\newcommand\bb[1] {\mbox{\boldmath{$#1$}}}
\newcommand{\Alfven}{Alfv\'{e}n~}
\newcommand\tomega{{\tilde\omega}}

\title[Low-$T/|W|$ instabilities in proto-neutron stars]
{Low-$T/|W|$ instabilities in differentially rotating proto-neutron stars with magnetic fields}

\author[W.~Fu \& D.~Lai]{Wen Fu$^1$\thanks{Email: wenfu@astro.cornell.edu (WF);
dong@astro.cornell.edu (DL)} and Dong Lai$^1$\footnotemark[1]\\
$^1$Department of Astronomy, Cornell University, Ithaca, NY 14853}

\pagerange{\pageref{firstpage}--\pageref{lastpage}} \pubyear{2010}
\begin{document}
\label{firstpage}
\maketitle

\begin{abstract}
Recent hydrodynamical simulations have shown that differentially
rotating neutron stars formed in core-collapse supernovae may develop
global non-axisymmetric instabilities even when $T/|W|$ (the ratio of
the rotational kinetic energy $T$ to the gravitational potential
energy $|W|$) is relatively small (less than 0.1). Such low-$T/|W|$
instability can give rise to efficient gravitational wave emission
from the proto-neutron star. We investigate how this instability is
affected by magnetic fields using a cylindrical stellar model.  Wave
absorption at the corotation resonance plays an important role in
facilitating the hydrodynamic low-$T/|W|$ instability.  In the
presence of a toroidal magnetic field, the corotation resonance is
split into two magnetic resonances where wave absorptions take place.
We show that the toroidal magnetic field suppresses the low-$T/|W|$
instability when the total magnetic energy $W_{\rm B}$ is of order $0.2\,T$
or larger, corresponding to toroidal fields of a few $\times 10^{16}$~G
or stronger.  Although poloidal magnetic fields do not influence the
instability directly, they can affect the instability by generating
toroidal fields through linear winding of the initial poloidal field and
magneto-rotational instability. We show that an initial poloidal field
with strength as small as $10^{14}$~G may suppress the low-$T/|W|$
instability.

\end{abstract}

\begin{keywords}
gravitational waves -- hydrodynamics -- instabilities -- MHD -- stars: neutron -- stars: rotation
\end{keywords}

\section{Introduction}
Rotating neutron stars (NSs) formed in the core collapse of a massive star or the accretion-induced collapse of a white dwarf maybe subject to nonaxisymmetric instabilities (e.g, Andersson 2003; Stergioulas 2003; Ott 2009). The onset and development of these rotational instabilities are often parameterized by the ratio $\beta \equiv T/|W|$, where $T$ is the rotational kinetic energy and $W$ the gravitational potential energy of the star.  In particular, the dynamical bar-mode ($m=2$) instability sets in when $\beta \gtrsim 0.27$ and grows on the dynamical time-scale. This critical $\beta$, originally derived for incompressible Maclaurin spheroid in Newtonian gravity (Chandrasekhar 1969), is relatively insensitive to the stiffness of the equation of state as long as the degree of differential rotation is not too large (e.g., Toman et al. 1998; New, Centrella \& Tohline 2000; Liu \& Lindblom 2001), although simulations show that it tends to be reduced by general relativity effect as the compactness of the star $M/R$ increases (Shibata, Baumgarte \& Shapiro 2000; Saijo et al. 2001). There also exist secular instabilities, which are driven by some dissipative mechanisms, such as viscosity and gravitational radiation. In the latter case it is known as the Chandrasekhar-Friedman-Schutz instability (Chandrasekhar 1970; Friedman \& Schutz 1978). Although the threshold of the secular bar-mode instability ($\beta \gtrsim 0.14$) is easier to be satisfied than the dynamical bar-mode instability, it grows on a much longer time-scale due to its dissipative nature
(e.g., Lai \& Shapiro 1995; Andersson 2003).

The nonlinear development of the dynamical bar-mode instability has been extensively studied in a large number of numerical simulations (Tohline, Durisen \& McCollough 1985; Pickett, Durisen \& Davis 1996; Cazes \& Tohline 2000; Brown 2000; Liu 2002; Shibata \& Sekiguchi 2005; Camarda et al. 2009). In the early 2000, it was found that for stars with sufficiently large differential rotation, dynamical instability can develop at significantly lower $\beta$ than $0.27$ (Centrella et al. 2001), even for $\beta$ on the order of $0.01$ (Shibata, Karino \& Eriguchi 2002, 2003; Ott et al. 2005; Ou \& Tohline 2006; Saijo \& Yoshida 2006; Cedra-Duran, Quilis \& Font 2007; Corvino et al. 2010). These low-$T/|W|$ appear to have quite different physical origin from the canonical bar-mode instability (see below). Most importantly, recent 3D simulations of a large sample of rotational core-collapse models carried out by Dimmelmeier et al. (2008), which include a state-of-the-art treatment of the microphysics during collapse and the initial rotational profiles obtained from models of precollapse evolution of massive stellar cores, have shown that in many of the models, the proto-NSs exhibit sufficient differential rotation to be subject to the low-$T/|W|$ instability (see also Ott et al. 2007)
\footnote{By contrast, the threshold for the canonical bar-mode instability is never reached even when the precollapse core has a very large angular momentum, because in that case core bounce would occur at low densities.}.
Such proto-NSs would generate strong gravitational waves (GWs), much stronger than a non-rotating core-collapse would produce, significantly increasing the possibility of detecting GWs from extra-galactic core-collapse supernovae by LIGO and other ground-based GW detectors (Ott 2009). We note that our current understanding of the angular momentum evolution of pre-supernova stars is uncertain, so one cannot predict the rotation profile of the collapsing core with great confidence (Heger, Woosley \& Spruit 2005). Therefore the detection (or non-detection) of the rotational signature of proto-NSs by GW detectors (such as Advanced LIGO) may provide valuable information on massive star evolution and the mechanism of core-collapse supernova explosion.

Despite clear numerical evidence for their existence, the physical origin of the low -$T/|W|$ instabilities remains unclear. It has been suggested (Watts, Andersson \& Jones 2005; Saijo \& Yoshida 2006) that the instabilities are associated with the existence of corotation resonance (where the wave pattern speed equals the background fluid rotation rate) inside the star and are thus likely to be a subclass of shear instabilities. Corotation resonance has long been known to be the key ingredient for some instabilities in other astrophysical fluid systems, such as the Papaloizou-Pringle instability for accretion torii (Papaloizou \& Pringle 1984; Fu \& Lai 2010b) and the corotational instability for thin accretion discs (Narayan, Goldreich \& Goodman 1987; Tsang \& Lai 2008, 2009; Lai \& Tsang 2009; Fu \& Lai 2010a). In addition, numerical calculations by Ou \& Tohline (2006) suggested that the presence of a local minimum in the radial vortensity profile of the star is also needed to amplify the mode (see also Corvino et al. 2010).

An important issue concerning the low-$T/|W|$ instability is the effects of magnetic fields. Proto-NSs are expected to contain appreciable magnetic fields. In particular, large toroidal fields can be generated from twisting relatively weak poloidal fields by differential rotation or from magneto-rotatioinal instabilities (e.g., Balbus \& Hawley 1998; Akiyama et al. 2003; Obergaulinger et al. 2009). While magnetic fields have a negligible effect on the high $T/|W|$ instability (Camarda et al. 2009), it is not clear whether low-$T/|W|$ can survive in the presence of B fields. Indeed, our previous work on magnetized discs showed that even a weak magnetic field can change the structure of corotation resonance significantly (Fu \& Lai 2010a). In this paper, as a first step of clarifying this issue, we carry out eigenvalue calculation of the effects of purely toroidal B fields on low $T/|W|$ instability by employing a cylindrical stellar model. This paper is the third in our series devoted to study the effects of magnetic fields on the global instabilities of various astrophysical flows, with the previous two focusing on black-hole accretion discs (Fu \& Lai 2010a) and accretion toii (Fu \& Lai 2010b), respectively. 

Our paper is organized as follows. In section 2, we describe the
equilibrium model for our rotating magnetized star. In section 3,
the linearized perturbation equations are presented and boundary conditions derived. In section 4 we present results from our numerical calculations. Final summary and discussion of our results are given in section 6.

\section{Equilibrium Model of A Magnetized Rotating Cylinder}
We consider a rotating star with purely toroidal magnetic fields and assume a polytropic equation of state
\be
P=K\rho^{\Gamma}=K\rho^{1+1/N},
\label{eq:poly}
\ee
where $P$ and $\rho$ are the gas pressure and density, $K$, $\Gamma$ and $N$ are constants. Although hydromagnetic stellar equilibrium models can be constructed easily using the HSCF method (Hachisu 1986; Tomimura \& Eriguchi 2005; see Lander \& Jones 2009 for recent works on uniformly rotating stars), linear eigenvalue analysis for such models is difficult. Thus we follow the setup in Saijo \& Yoshida (2006) by treating the star as an infinite cylinder. We adopt the cylindrical coordinates ($r$, $\phi$, $z$). All the background variables are assumed to be functions of cylindrical radius $r$ only. The equilibrium state of the cylinder is determined by force balance equation in the radial direction
\be
\frac{1}{\rho}\frac{d P}{dr}=-\frac{d\Phi}{dr}+r\Omega^2-\frac{1}{\rho}
\frac{dP_{\rm m}}{dr}-\frac{B_{\phi}^2}{4\pi \rho r},
\label{eq:equili}
\ee
where $\Omega$ is the flow rotation rate, $B_{\phi}$ is the toroidal magnetic field strength, $P_{\rm m}=B_{\phi}^2/8\pi$ is the magnetic pressure, and $\Phi$ is the Newtonian gravitational potential which relates to density $\rho$ via Poisson's equation
\be
\nabla^2\Phi=4\pi G\rho.
\label{eq:poisson}
\ee
Eliminating $\Phi$ from Eqs. (\ref{eq:equili}) and (\ref{eq:poisson}) yields
\be
\frac{d}{dr}\left(\frac{r}{\rho}\frac{d P}{dr}\right)
=-4\pi G\rho r+\frac{d}{dr}(r^2\Omega^2)-\frac{d}{dr}\left(\frac{r}{\rho}
\frac{dP_{\rm m}}{dr}\right)-\frac{d}{dr}\left(\frac{B_{\phi}^2}{4\pi \rho}\right).
\label{eq:Emden1}
\ee

For numerical convenience, we nondimensionalize variables as follows
\be
\rho=\hat{\rho}\rho_c=\theta^N \rho_c,
\ee
\be
r=\hat{r}\sqrt{\frac{(N+1)K\rho_c^{1/N-1}}{4\pi G}},
\ee
\be
\Omega=\hat{\Omega}\sqrt{4\pi G \rho_c},
\ee
\be
B_\phi=\hat{B_\phi}\sqrt{4\pi(N+1)K\rho_c^{1+1/N}},
\ee
where $\rho_c$ is the central density and the hatted variables denote dimensionless quantities. Similar dimensionless variables for other quantities can be constructed from the list above.
We follow Saijo \& Yoshida (2006) to adopt the following rotation profile:
\be
\hat{\Omega}=\frac{C}{\hat{r}^2+A},
\ee
where $A$ and $C$ are constants. The toroidal B field profile we employ is
\be
\hat{B_{\phi}}=b\hat{r}(\hat{R}-\hat{r})
\label{eq:bphi}
\ee
where $\hat{R}$ denotes the dimensionless boundary of the cylinder and the constant $b$ specifies the field strength. For simplicity, we will omit the hats on all variables hereafter unless otherwise noted. The above profile implies that $B_{\phi}=0$ at both the center and the surface of the star. For small $r$, we have $B_{\phi} \simeq brR$, implying a constant axial current for $r \rightarrow 0$. Eq. (\ref{eq:Emden1}) in dimensionless form now reads
\be
\frac{d^2 \theta}{dr^2}+\left[\frac{1}{r}-N\theta^{-N-1}b^2r(R-r)(R-2r)\right]\frac{d\theta}{dr}
+\theta^{N}+\left[4b^2(R-r)(R-2r)-b^2r(3R-4r)\right]\theta^{-N}=2\Omega\frac{d}{dr}(r\Omega).
\label{eq:Emden2}
\ee
For an nonmagnetized star, Eq.~(\ref{eq:Emden2}) reduces to Eq.~(3.2) in Saijo \& Yoshida (2006). In the limit of zero rotation, it recovers the
well-known Lane-Emden equation in cylindrical geometry.

\begin{figure}
\begin{center}
\includegraphics[scale=0.55]{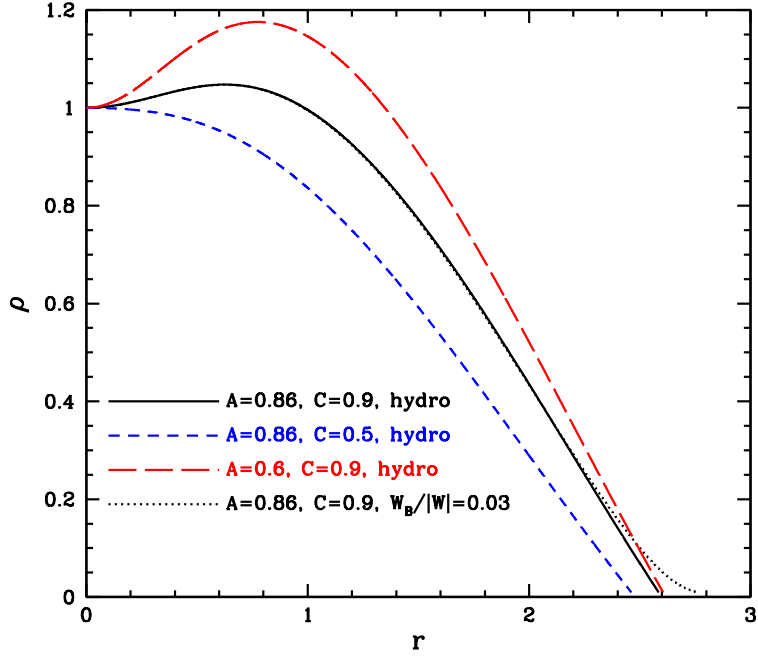}
\caption{Density profiles for hydrodynamic and hydromagnetic equilibria. The polytropic index is $N=1$. The solid, short-dashed and long-dashed lines represent different rotation profiles for non-magnetic models, while the dotted line is for a magnetic model with $W_{\rm B}/|W| =0.03$.}
\label{fig:theta}
\end{center}
\end{figure}

In the limit of $B_{\phi}=0$, we can simply integrate Eq.~(\ref{eq:Emden2}) starting from $r=0$ using boundary condition that $\theta=1$ and $\theta'=0$ to a point where $\theta$ goes to zero, which defines the cylinder surface $R$. The hydrodynamic equilibrium can thus be easily constructed. When $B_{\phi}$ is non-zero, we choose an initial guess for the surface radius $R$ based on results for the equivalent hydrodynamic model and integrate Eq.~(\ref{eq:Emden2}) imposing the same boundary condition at the center. We stop the integration at $r=R$ to check the value of $\theta$. We then adjust our guess for $R$ and go through the same process, until $\theta|_{r=R}$ comes close enough to $0$. For a given equilibrium state, the rotational kinetic energy $T$, gravitational potential energy $W$ and magnetic energy $W_{\rm B}$ of the cylinder have the following form
\[
T=\frac{1}{2}\int \rho r^2\Omega^2dV
=\int_0^{R} \theta^N \Omega^2 r^3 dr,
\]
\[
W=-\int \rho r\frac{d\Phi}{dr}dV
=\left(\int_0^{R}\theta^N rdr\right)^2,
\]
\be
W_{\rm B}=\int \frac{B_\phi^2}{8\pi}dV=\frac{R^6b^2}{60},
\ee
where all the variables are dimensionless and the corresponding physical unit for energy is
$(N+1)^2K^2\rho_c^{2/N}/4$. Examples of the equilibrium density profile are given in Fig.~\ref{fig:theta}. We see that the density profile is not always monotonic: for large $C$ and small $A$ (i.e, large rotation rate and large degree of differential rotation), the density maximum is off-centered.

\section{Linear Perturbation Analysis}
\subsection{Perturbation equations}
The cylindrical flow we are considering satisfies the usual ideal MHD equations
\be
{\partial{\rho} \over \partial t}
+\nabla\cdot(\rho \bb{v})=0,
\label{eq:mhd1}
\ee
\be
{\partial{\bb{v}} \over \partial
t}+({\bb{v}}\cdot\nabla){\bb{v}}
=-\frac{1}{\rho}\nabla \Pi-{\nabla \Phi}
+\frac{1}{4\pi\rho}(\bb{B}\cdot\nabla)\bb{B},
\label{eq:mhd2}
\ee
\be
{{\partial {\bb{B}}} \over \partial t}=\nabla\times({\bb{v}}
\times{\bb{B}}),
\label{eq:mhd3}
\ee
\be
\nabla \cdot \bb{B}=0,
\label{eq:mhd4}
\ee
\be
\nabla^2\Phi=4\pi G\rho,
\label{eq:mhd5}
\ee
where $\Pi=P+P_{\rm m}$ is the total pressure. We apply linear perturbations to the above equations by assuming the perturbation of any physical variable $f$ to have the form $\delta f \propto {\rm e}^{{\rm i}m\phi-{\rm i}\omega t}$ with $m$ being the azimuthal mode number and $\omega$ the wave frequency. The resulting linearized perturbation equations contain variables $\delta \bb{v}$, $\delta \rho$, $\delta \Pi$, $\delta \Phi$ and $\delta \bb{B}$. To simply the algebra, we define a new variable
\[
\delta h=\frac{\delta \Pi}{\rho}=\frac{\delta P}{\rho}+\frac{\bb{B}\cdot \delta \bb{B}}{4\pi\rho}.
\]
Using $\Delta \bb{v}=\delta
\bb{v}+\bb{\xi}\cdot\nabla\bb{v}=d\bb{\xi}/dt
=-{\rm i}\omega\bb{\xi}+(\bb{v}\cdot \nabla)\bb{\xi}=-{\rm i}\omega\bb{\xi}+\Omega \partial \bb{\xi}/\partial \phi$, we find that the
Eulerian perturbation $\delta \bb{v}$ is related to the Lagrangian
displacement vector $\bb{\xi}$ by $\delta
\bb{v}=-\rm{i}\tomega\bb{\xi}-r\Omega'\xi_r \bb{\hat{\phi}}$ (prime denotes
radial derivative). In terms of $\xi_r$, $\delta h$ and $\delta \Phi$, the MHD perturbation equations (in dimensionless form) can be cast into four first-order differential equations:
\be
\frac{d\xi_r}{dr}=A_{11}\xi_r+A_{12}\delta h+A_{13}\delta \Phi+A_{14}\frac{d\delta \Phi}{dr},
\label{eq:ode1}
\ee
\be
\frac{d\delta h}{dr}=A_{21}\xi_r+A_{22}\delta h+A_{23}\delta \Phi+A_{24}\frac{d\delta \Phi}{dr},
\label{eq:ode2}
\ee
\be
\frac{d\delta \Phi}{dr}=A_{31}\xi_r+A_{32}\delta h+A_{33}\delta \Phi+A_{34}\frac{d\delta \Phi}{dr},
\label{eq:ode3}
\ee
\be
\frac{d}{dr}\left(\frac{d\delta \Phi}{dr}\right)=A_{41}\xi_r+A_{42}\delta h+A_{43}\delta \Phi+A_{44}\frac{d\delta \Phi}{dr},
\label{eq:ode4}
\ee
where
\be
A_{11}=\frac{r\tomega^2\left[(\omega_{A\phi}^2-\Omega^2)\tomega^2
+\omega_{A\phi}^2\omega^2\right]}{(c_s^2+v_{A\phi}^2)(\tomega^2-m^2\omega_{A\phi}^2)
(\tomega^2-\omega_s^2)}
+\frac{g\tomega^2}{(c_s^2+v_{A\phi}^2)(\tomega^2-\omega_s^2)}
-\frac{\tomega^2+2m\tomega\Omega+m^2\omega_{A\phi}^2}{r(\tomega^2-m^2\omega_{A\phi}^2)},
\label{eq:a11}
\ee
\be
A_{12}=-\frac{\tomega^4}{(c_s^2+v_{A\phi}^2)(\tomega^2-m^2\omega_{A\phi}^2)
(\tomega^2-\omega_s^2)}+\frac{m^2}{r^2(\tomega^2-m^2\omega_{A\phi}^2)},
\label{eq:a12}
\ee
\be
A_{13}=\frac{m^2}{r^2\tomega^2},
\ee
\be
A_{14}=0,
\ee
\[
A_{21}=\tomega^2-m^2\omega_{A\phi}^2-\frac{4(m\omega_{A\phi}^2+\tomega\Omega)^2}
{\tomega^2-m^2\omega_{A\phi}^2}+r\frac{d}{dr}(\omega_{A\phi}^2-\Omega^2)
+(\omega_{A\phi}^2-\Omega^2)\frac{r}{\rho}\frac{d\rho}{dr}
+\frac{g}{\rho}\frac{d\rho}{dr}
\]
\be
~~~~~~~+\frac{1}{(c_s^2+v_{A\phi}^2)(\tomega^2-m^2\omega_{A\phi}^2)(\tomega^2-\omega_s^2)}
\left\{r\left[(\omega_{A\phi}^2-\Omega^2)\tomega^2+\omega_{A\phi}^2\omega^2\right]
+g(\tomega^2-m^2\omega_{A\phi}^2)\right\}^2,
\label{eq:a21}
\ee
\be
A_{22}=-\frac{r\tomega^2\left[(\omega_{A\phi}^2-\Omega^2)\tomega^2
+\omega_{A\phi}^2\omega^2\right]}{(c_s^2+v_{A\phi}^2)(\tomega^2-m^2\omega_{A\phi}^2)
(\tomega^2-\omega_s^2)}
-\frac{g\tomega^2}{(c_s^2+v_{A\phi}^2)(\tomega^2-\omega_s^2)}
+\frac{2m(m\omega_{A\phi}^2+\tomega\Omega)}{r(\tomega^2-m^2\omega_{A\phi}^2)}
-\frac{1}{\rho}\frac{d\rho}{dr}.
\label{eq:a22}
\ee
\be
A_{23}=\frac{2m\Omega}{r\tomega},
\ee
\be
A_{24}=-1,
\ee
\be
A_{31}=A_{32}=A_{33}=0,
\ee
\be
A_{34}=1
\ee
\be
A_{41}=-\frac{\rho}{r}-\frac{d\rho}{dr}
+\rho\left(\frac{m^2\omega_{A\phi}^2}{\tomega^2}-1\right)A_{11}
-\rho\frac{m^2\omega_{A\phi}^2}{r\tomega^2}-\frac{2\rho m\Omega}{r\tomega},
\ee
\be
A_{42}=\rho\left(\frac{m^2\omega_{A\phi}^2}{\tomega^2}-1\right)A_{12}
+\rho\frac{m^2}{r^2\tomega^2},
\ee
\be
A_{43}=\rho \frac{m^4 \omega_{A\phi}^2}{r^2\tomega^4}+\frac{m^2}{r^2},
\ee
\be
A_{44}=-\frac{1}{r}.
\ee
In the above expressions, $\tomega=\omega-m\Omega$ is the wave frequency in the co-rotating frame, $\rho=\theta^{N}$ is the dimensionless density, $c_s=\sqrt{dP/d\rho}=\sqrt{\theta/N}$ is the dimensionless sound speed,
\be
v_{A\phi}=\sqrt{\frac{B_{\phi}^2}{4\pi \rho}}=br(R-r)\theta^{-N/2}
\ee is the toroidal \Alfven velocity,
$\omega_{A\phi}=v_{A\phi}/r=b(R-r)\theta^{-N/2}$ is the toroidal \Alfven frequency,
\be
\omega_s=\sqrt{\frac{c_s^2}{c_s^2+v_{A\phi}^2}m^2\omega_{A\phi}^2}
\label{eq:omegas}
\ee
is the slow magnetosonic wave frequency for $\bb{k}=(m/r)\bb{\hat{\phi}}$,
and
\be
g=\frac{d\Phi}{dr}=r\Omega^2-\frac{d\theta}{dr}-\left[b^2r(R-r)(2R-3r)\right]\theta^{-N}
\ee
is the gravitational acceleration in radial direction.

\subsection{Boundary conditions}
To solve Eqs.~(\ref{eq:ode1})-(\ref{eq:ode4}) as an eigenvalue problem, we need four boundary conditions. The outer boundary conditions are straightforward. From the perturbed Poisson equation, we know the perturbed potential outside the star scales as $\delta \Phi \propto r^{-m}$. By requiring this potential to match smoothly with the potential inside, we obtain our first outer boundary condition:
\be
\frac{d\delta \Phi}{dr}+\frac{m}{r}\delta \Phi=0~~~\mbox{at}~~~r=R.
\ee
Requiring the Lagrangian pressure perturbation to vanish at the stellar surface yields
\be
\delta h+\frac{d\theta}{dr}\xi_r=0~~~\mbox{at}~~~r=R.
\label{eq:obc1}
\ee
The inner boundary conditions are more involved. As $r\rightarrow 0$, we observe that $g\rightarrow 0$, $\Omega\rightarrow \mbox{constant}$, $\rho \rightarrow \mbox{constant}$, $\omega_{A\phi}\propto B_{\phi}/r \rightarrow \mbox{constant}$ and $\delta \rho$ is finite. Thus, near the center of the star, Eqs.~(\ref{eq:ode1})-(\ref{eq:ode4}) can be simplified as
\be
\frac{d\xi_r}{dr}=-\frac{X+mY}{X}\frac{\xi_r}{r}+\frac{m^2}{X}\frac{\delta h}{r^2}
+\frac{m^2}{\tomega^2}\frac{\delta \Phi}{r^2}
\label{eq:sode1}
\ee
\be
\frac{d\delta h}{dr}=\frac{X^2-Y^2}{X}\xi_r+\frac{mY}{X}\frac{\delta h}{r}
+\frac{2m\Omega}{\tomega}\frac{\delta \Phi}{r}-\frac{d\delta \Phi}{dr},
\label{eq:sode2}
\ee
\be
\frac{1}{r^2}\frac{d}{dr}\left(r^2\frac{d\delta \Phi}{dr}\right)-\frac{m^2}{r^2}\delta \Phi=0,
\label{eq:sode3}
\ee
where,
\be
X=\tomega^2-m^2\omega_{A\phi}^2,~~~Y=2(m\omega_{A\phi}^2+\tomega\Omega).
\ee
Since perturbation equations are linear, we simply take the solution of Eq. (\ref{eq:sode3}) to be
\be
\delta \Phi=r^m, ~~~\mbox{at}~~~r\sim0.
\ee
The perturbations $\xi_r$ and $\delta h$ generally take the form
\be
\xi_r=C_1r^{m-1}+C_2r^{m-1}\ln r,~~~\delta h=C_3r^m+C_4r^m\ln r,
\label{eq:solution}
\ee
where $C_1,~C_2,~C_3,~C_4$ are constants so that perturbations remain regular at the center. Eq.~(\ref{eq:solution}) represents the leading terms of the Frobenius expansions of these functions.

Substituting solution (\ref{eq:solution}) into Eqs.~(\ref{eq:sode1}) and (\ref{eq:sode2}) leads to two equations which have the structure $a_1+a_2\ln r =0$,
where $a_1,~a_2$ are constants that depend on the values of $m,~\Omega,~\tomega,~\omega_{A\phi},~C_1,~C_2,~C_3,~C_4$ near the center. Since these equations should be satisfied everywhere around the center, we demand $a_1=a_2=0$. This yields
\be
C_2+\frac{m(X+Y)}{X}C_1=\frac{m^2}{X}C_3+\frac{m^2}{\tomega^2},
\ee
\be
mC_4=(X+Y)C_2,
\ee
\be
C_4+\frac{m(X-Y)}{X}C_3=\frac{X^2-Y^2}{X}C_1+m\frac{2\Omega-\tomega}{\tomega},
\ee
from which we can determine three constants
\be
C_2=-\frac{m^3(m+2)}{2X}\omega_{A\phi}^2,
\ee
\be
C_4=-\frac{m^2(m+2)}{2}\frac{X+Y}{X}\omega_{A\phi}^2,
\ee
\be
C_1=\frac{m}{X+Y}C_3+\frac{m^2(m+2)}{2(X+Y)}\omega_{A\phi}^2+\frac{m}{\tomega^2}
\frac{X}{X+Y},
\label{eq:AC}
\ee
once we specify $C_3$. When we solve the eigenvalue problem, $C_3$ will be determined together with the eigenfrequency $\omega$. We see that since for the specific $B_{\phi}$ profile we are considering (see Eq.~[\ref{eq:bphi}]), $\omega_{A\phi}$ remains approximately constant near the center, $C_2$ and $C_4$ are both finite. Therefore the logarithmic parts in solution (\ref{eq:solution}) cannot be neglected.

In the hydrodynamic limit ($B_{\phi}=0,~\omega_{A\phi}=0$), we have $C_2=C_4=0$ so that the solutions of $\xi_r$ and $\delta h$ take a purely power-law form, and   Eq.~(\ref{eq:AC}) reduces to
\be
C_1=\frac{m}{\tomega(\tomega+2\Omega)}(C_3+1),
\ee
which is equivalent to
\be
\xi_r=\frac{m}{r\tomega(\tomega+2\Omega)}(\delta h+\delta \Phi)~~~\mbox{at}~~~r\sim 0.
\label{eq:hydrobc}
\ee
Again, constant $C_3=\delta h/\delta \Phi$ will be determined as a part of the eigenvalue problem. Clearly, for the magnetic cases where $d\ln B_{\phi}/d\ln r> 1~\mbox{at}~r\sim 0$ so that $\omega_{A\phi}\rightarrow 0$ as $r\rightarrow 0$, the above hydrodynamic boundary condition is also valid.

\subsection{Cowling approximation}
In the Cowling approximation, we
neglect the gravitational potential perturbation $\delta \Phi$. The perturbation equations then become
\be
\frac{d\xi_r}{dr}=A_{11}\xi_r+A_{12}\delta h,
\label{eq:code1}
\ee
\be
\frac{d\delta h}{dr}=A_{21}\xi_r+A_{22}\delta h,
\label{eq:code2}
\ee
with the four coefficients given by the same equations as before.
Similarly, for $r\rightarrow 0$ the simplified version of Eqs.~(\ref{eq:sode1}) and (\ref{eq:sode2}) are
\be
\frac{d\xi_r}{dr}=-\frac{X+mY}{X}\frac{\xi_r}{r}+\frac{m^2}{X}\frac{\delta h}{r^2}
\label{eq:sode4}
\ee
\be
\frac{d\delta h}{dr}=\frac{X^2-Y^2}{X}\xi_r+\frac{mY}{X}\frac{\delta h}{r}.
\label{eq:sode5}
\ee
The outer boundary condition in this case is again given by Eq.~(\ref{eq:obc1})
\be
\delta h+\frac{d\theta}{dr}\xi_r=0~~~\mbox{at}~~~r=R.
\label{eq:bc1}
\ee
The inner boundary condition can be obtained by substituting the power-law solutions
$\xi_r \propto r^{m-1}$ and $\delta h \propto r^m$ into Eqs.~(\ref{eq:sode4}) and (\ref{eq:sode5}), giving
\be
\xi_r=\frac{m}{r[\tomega^2+2\tomega\Omega-m(m-2)\omega_{A\phi}^2]}\delta h.
\label{eq:bc2}
\ee
Note that for either $m=2$ perturbations or a unmagnetized flow, the above inner B.C. reduces to the same form
\be
\xi_r=\frac{m}{r\tomega(\tomega+2\Omega)}\delta h.
\ee

\begin{figure}
\begin{center}
\includegraphics[scale=0.5]{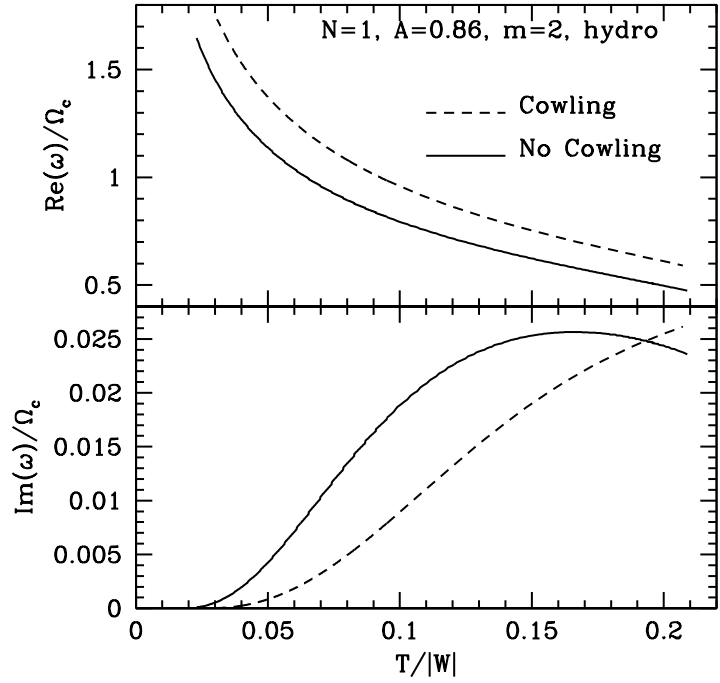}
\caption{The $m=2$ mode frequency as a function of $T/|W|$ with and with out Cowling approximation. The upper and bottom panels show the real and imaginary parts of the frequency, respectively, with $\Omega_{\rm c}$ being the rotation frequency at the center. The star has no magnetic field and the polytropic index is $N=1$.}
\label{fig:cowling}
\end{center}
\end{figure}

\begin{figure}
\begin{center}
\includegraphics[scale=0.5]{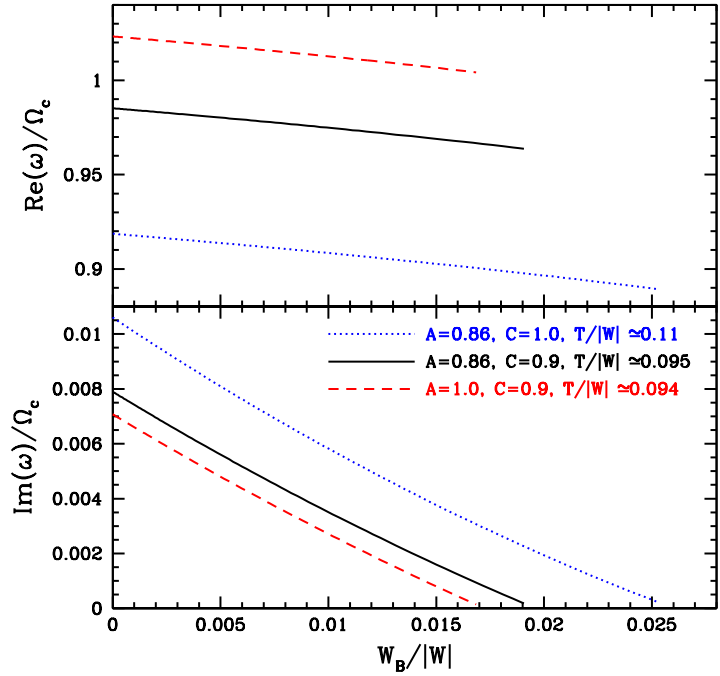}
\caption{The $m=2$ mode frequency as a function of $W_{\rm B}/|W|$ for stellar models with different rotation profiles (thus different $T/|W|$'s). The upper and bottom panels show the real and imaginary parts of the frequency, respectively, with $\Omega_{\rm c}$ being the rotation frequency at the center. The other parameters are the same as in Fig.~\ref{fig:cowling}.}
\label{fig:omega}
\end{center}
\end{figure}

\begin{figure}
\begin{center}
\includegraphics[scale=0.5]{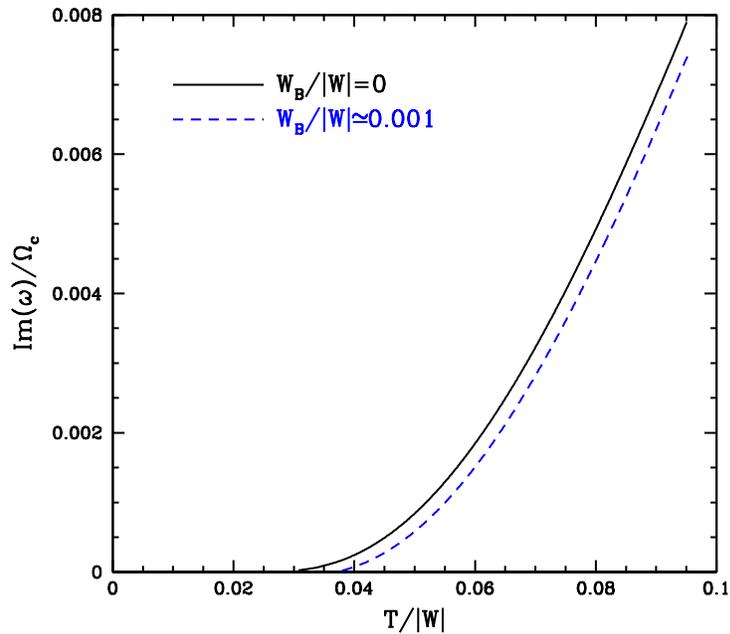}
\caption{The $m=2$ mode growth rate as a function of $T/|W|$ with and without toroidal magnetic field. The other parameters are $A=0.86$ and $N=1$.}
\label{fig:onset}
\end{center}
\end{figure}

\begin{figure}
\begin{center}$
\begin{array}{cc}
\includegraphics[scale=0.45]{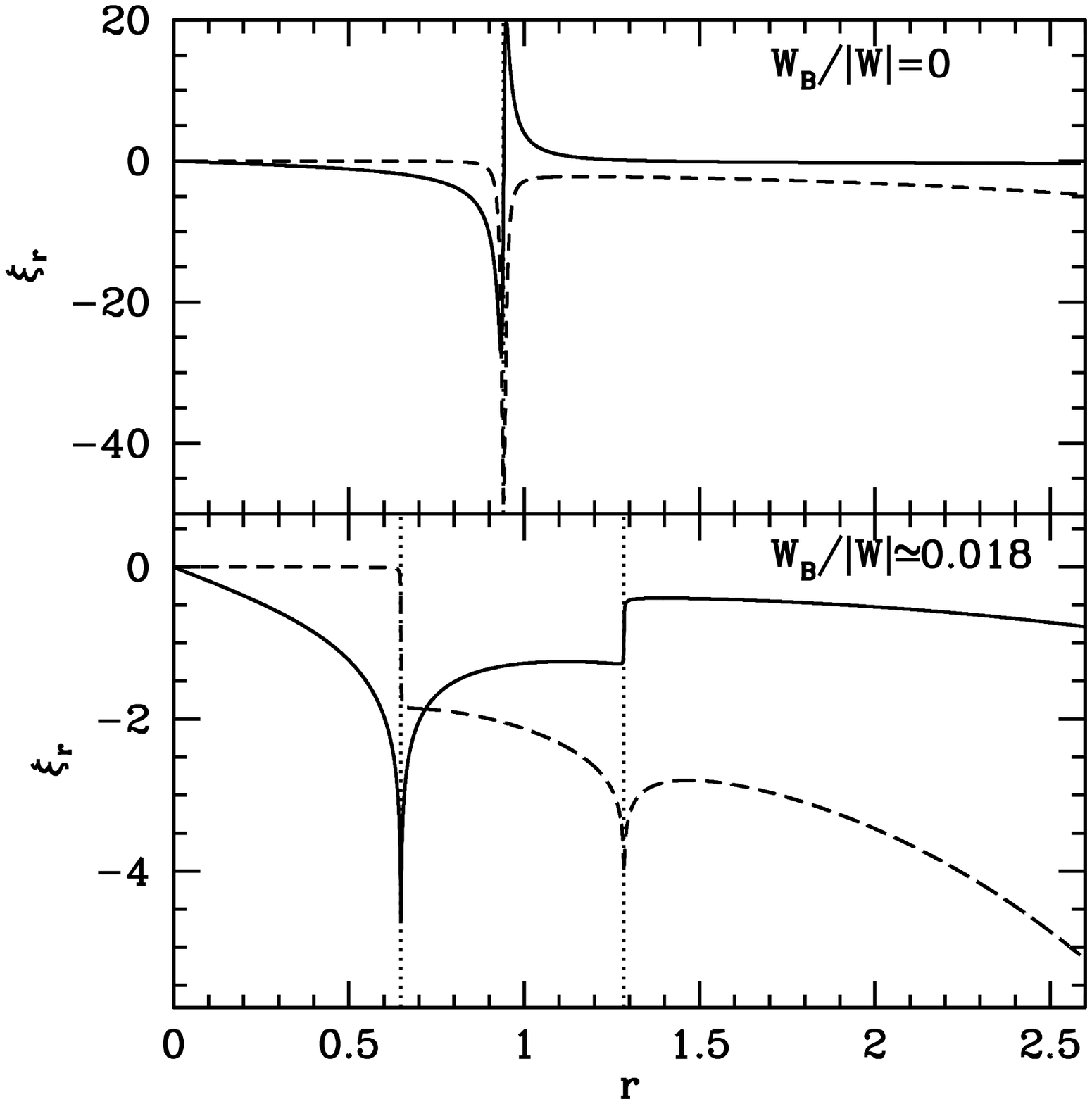} &
\includegraphics[scale=0.45]{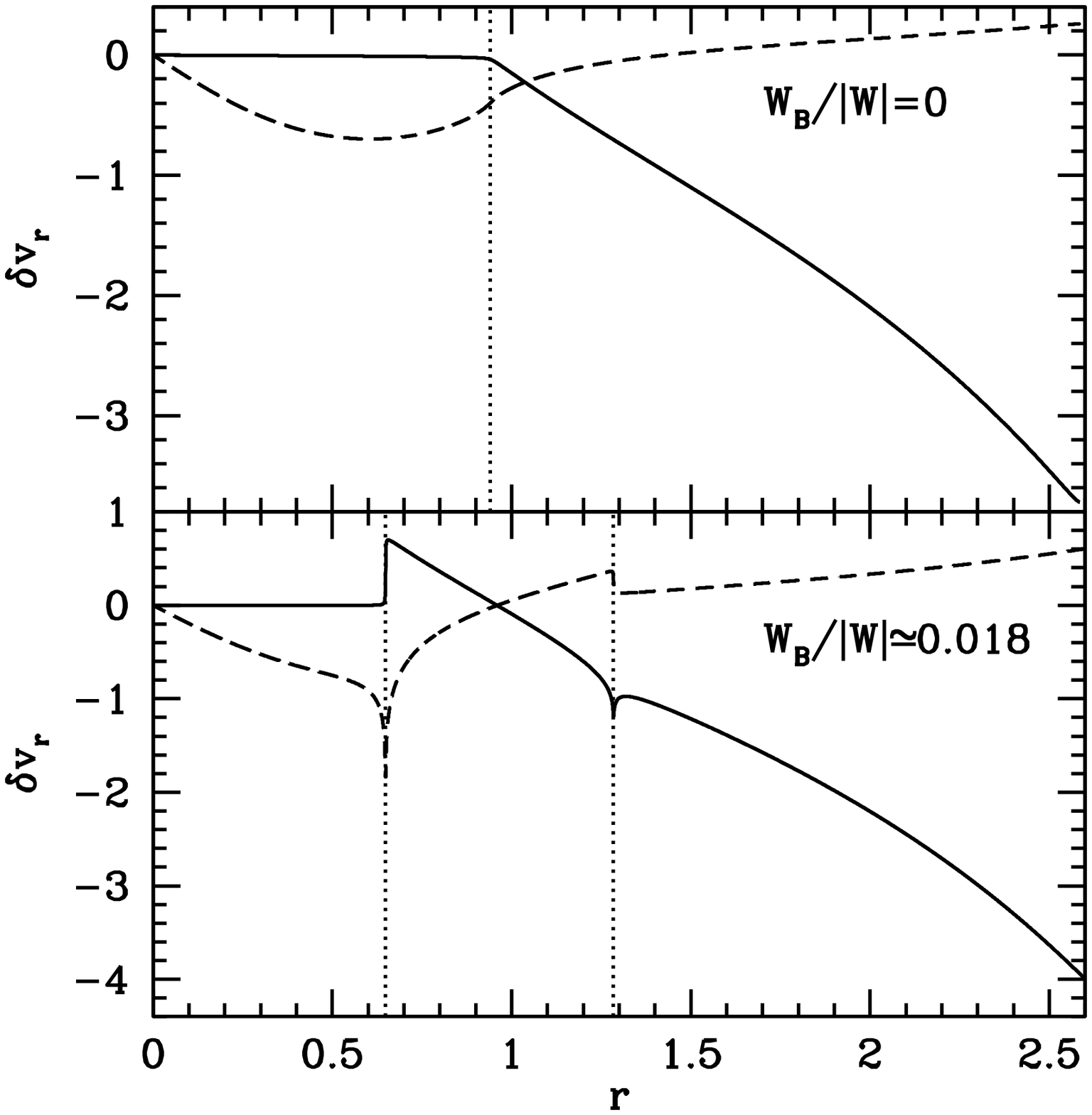}
\end{array}$
\caption{Example wavefunctions of unstable low-$T/|W|$ ($\simeq 0.095$) mode with $A=0.86$, $C=0.9$, $m=2$ and $N=1$. The left column shows the radial displacement as a function of radius whereas the right column shows the radial velocity perturbation, with the solid and short-dashed lines representing the real and imaginary parts, respectively. The upper and lower panels are for nonmagnetic and magnetic stellar models, respectively. The dotted lines indicate the location of the corotation resonance (in the nonmagnetic case) or slow magnetosonic resonances (in the magnetic case). The vertical scales of the wavefunctions are arbitrary. }
\label{fig:wave}
\end{center}
\end{figure}

\begin{figure}
\begin{center}
\includegraphics[scale=0.5]{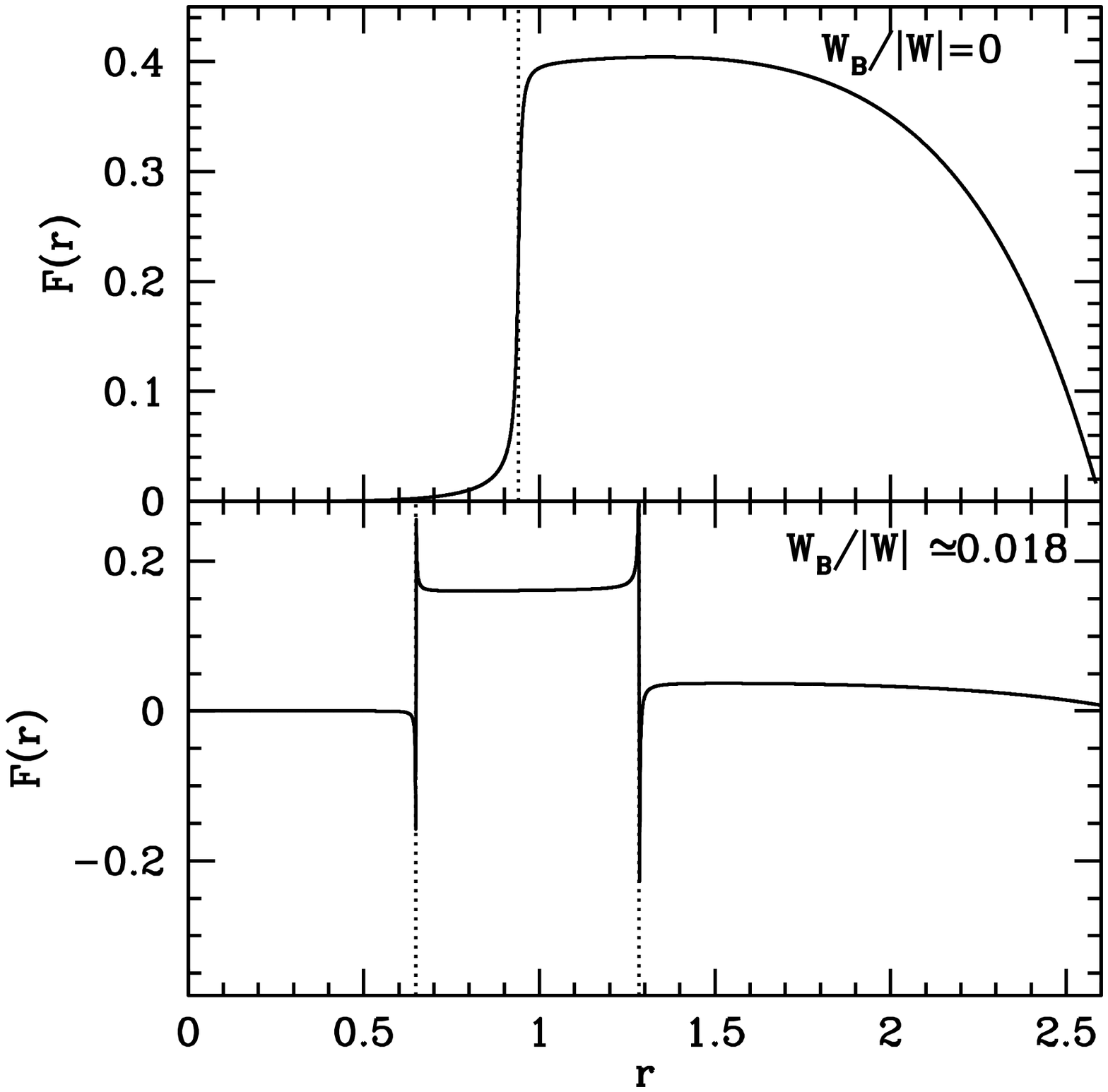}
\caption{Angular momentum carried by the wave as a function of $r$. The model parameters are the same as in Fig.~\ref{fig:wave}. The upper and lower panels show the nonmagnetic and magnetic models, respectively. The locations of the corotation resonance and slow resonances are indicated by the vertical dotted lines. }
\label{fig:flux}
\end{center}
\end{figure}

\section{Numerical Results}
For most part of this section, we will employ the standard shooting method (Press et al 1992) to solve the two ODEs, Eqs. (\ref{eq:code1}) and (\ref{eq:code2}), subject to boundary conditions (\ref{eq:bc1}) and (\ref{eq:bc2}). We focus on the effects of toroidal magnetic fields on the low-$T/|W|$ instability previously found for purely hydrodynamic stars.

Before moving on to our main results, let us first examine the validity of Cowling approximation for low-$T/|W|$ instability. To this end, we compare the eigenfrequency calculation with and without Cowling approximation. In Fig.~\ref{fig:cowling}, we fix one of the rotation parameters, $A$, while change the other, $C$, to obtain different values of $T/|W|$. We see that for the whole range of $T/|W|$ considered, the real part of the mode frequency does not show much difference between those two cases. The bottom panel of Fig.~\ref{fig:cowling} shows that, for the low $T/|W|$ range ($\lesssim 0.1$), the mode growth rate exhibits qualitatively similar behavior with an approximate factor of 2 difference between the two cases. For relatively large $T/|W|$ ($\gtrsim 0.2$), the growth rate no longer follows the similar trend when $T/|W|$ increases. Shibata et al. (2002) also found from their hydrodynamic simulation of a similar stellar model that the mode growth rate declines beyond certain $T/|W|$. The solid lines (``no Cowling'') in our Fig.~\ref{fig:cowling} agree well with the results depicted in Fig. 4 of Shibata et al.. With Cowling approximation, we find that the growth always increases with increasing $T/|W|$. Overall, Fig.~\ref{fig:cowling} shows that using the Cowling approximation captures the essential feature of the low-$T/|W|$ instability, especially when $T/|W|$ is not much larger than the threshold.

Figs.~\ref{fig:omega} and \ref{fig:onset} contain the most important results of this paper. In Fig.~\ref{fig:omega}, we plot the eigenfrequency of the $m=2$ mode as a function of $W_{\rm B}/|W|$ (the ratio of magnetic energy to gravitational energy) for different rotation profiles. Note that as we change the magnetic field strength, the equilibrium structure, therefore $T/|W|$ will also change. However, for the range of $W_{\rm B}/|W|$ we considered, the modification to the equilibrium structure is so small (see dotted line in Fig.~\ref{fig:theta} for the small modification) that $T/|W|$ is approximately a constant along the three curves. Fig.~\ref{fig:omega} demonstrates that the low-$T/|W|$ instability can be suppressed by the toroidal magnetic field. The point where the mode growth is completely suppressed corresponds to $W_{\rm B}/|W| \sim 0.2 ~T/|W|$. Fig.~\ref{fig:onset} shows the mode growth rate as a function of $T/|W|$ for stellar models with different $W_{\rm B}/|W|$. We see that the finite magnetic field shifts the curve towards larger $T/|W|$. In particular, the magnetic field increases the threshold for the instability from $T/|W| \simeq 0.03$ for the nonmagnetic model to $T/|W| \simeq 0.035$ for the $W_{\rm B}/|W| \simeq 0.001$ model. This finding can be easily understood: increasing rotation drives the instability, whereas magnetic field suppresses the instability. Therefore when a finite B field is included, in order to maintain the instability a larger rotation rate is needed to overcome the suppressing effect.

Fig.~\ref{fig:wave} depicts two example wavefunctions of the overstable low-$T/|W|$ mode. In the nonmagnetic case, the perturbation equations are singular at the corotation radius where $\tomega=\omega-m\Omega=0$. For low-$T/|W|$ modes, the corotation resonance lies inside the star, so both the radial displacement and the gradient of the radial velocity perturbation undergo large variations across the corotation resonance (see the upper panels). In the magnetic case, however, the corotation resonance is no longer a singularity. Instead, the perturbation equations are singular at two slow magnetosonic resonances where $\tomega=\pm \omega_s$, with $\omega_s$ given by Eq.~(\ref{eq:omegas})
\footnote{Note that $\tomega^2-m^2\omega_{A\phi}^2=0$ is not a singularity even though it appears to be a singular term similar to $\tomega^2-\omega_{s}^2$ in those coefficients of differential equations. This apparent singular term, one can show, will be canceled by some subtle mathematical manipulations. But this cancelation only works for the particular setup we consider here (pure toroidal B field, no vertical structure in perturbations). In general (i.e., with mixed B field or finite $k_z$), equations will be singular at both $\tomega^2=m^2\omega_{A\phi}^2$ and $\tomega^2=\omega_{s}^2$ (see Fu \& Lai 2010a).}.
Therefore, the wavefunctions exhibit sudden changes at these two particular locations (see the lower panels). This splitting of corotation resonance into two magnetic slow resonances can also be seen in the angular momentum flux. In Fig.~\ref{fig:flux}, we show the angular momentum carried by the wave across the star as a function of radius (see Fu \& Lai 2010a for the flux formula). In the upper panel (nonmagnetic case), we see that $F(r)$ experiences a sudden jump at the corotation resonance, whereas in the lower panel (magnetic case), two jumps occur at the two slow resonances and have different signs. This is similar to thin accretion discs studied in Fu \& Lai (2010a). However, since the outer boundary condition we employed here (free surface) is totally different from the one used in Fu \& Lai (2010a) (outgoing waves), we cannot directly relate the flux jump (or the net jump in the case of two resonances) to the magnitude of the growth rate. In any case, it is clear from Fig.~\ref{fig:flux} that the corotation resonance indeed plays an important role in driving the hydrodynamic low-$T/|W|$ instability and the toroidal magnetic field affects the instability by splitting the corotation resonance into two magnetic slow resonances. The property of the unstable mode in the presence of a magnetic field is determined by the combined effects from both slow resonances.

\section{Discussion}

Recent studies of rotating (but nonmagnetic) core-collapse supernovae
(e.g., Dimmelmeier et al.~2008)
have demonstrated that newly formed neutron stars can develop
nonaxisymmetric global instabilities with low $T/|W|$, and such instabilities
lead to significant gravitational wave emission.
In this paper, we have carried out the linear stability analysis of
magnetic, differentially rotating stars (modeled as a cylinder) to
examine how magnegic fields affect the low-$T/|W|$ rotational
instability. We show that the wave absorption at the corotation resonance
plays an important role in the instability. In the presence of a
toroidal magnetic field, the corotation resonance is split into two
magnetic resonances, where wave absorptions of opposite signs take
place. Our main result is that toroidal magnetic fields reduce the
growth rate of the low-$T/|W|$ instability and increase the threshold
$T/|W|$ value above which the instability occurs. To significantly
affect the instability, the required $W_{\rm B}/|W|$ (the ratio of the
magnetic energy $W_{\rm B}$ to the gravitational potential energy $|W|$)
should be of order $0.2\,T/|W|$ or larger (see Figs.~\ref{fig:omega}-\ref{fig:onset}).
As the
critical $T/|W|$ ranges from 0.01 to 0.1, the required $W_{\rm B}/|W|$
lies between 0.002 and 0.02. Using $|W|\sim (3/5)GM^2/R$ and $W_{\rm B}\sim
(B_\phi^2/8\pi)(4\pi R^3/3)$, we have
\be
{W_{\rm B}\over |W|}\sim {1\over 300}
\left({B_\phi\over 2\times 10^{16}\,{\rm G}}\right)^2
\left({R\over 20\,{\rm km}}\right)^4
\left({M\over 1.4M_\odot}\right)^{-2}.
\ee
Thus, only toroidal magnetic fields stronger than $2\times 10^{16}$~G can
significantly affect the low-$T/|W|$ instability.

In our simple (cylindrical) stellar model, poloidal (vertical)
magnetic fields do not directly affect the rotational instability
because the unstable modes do not have vertical structure (i.e., the
vertical wavenumber is zero). We believe that this also holds for
more realistic stellar models, although more investigations are needed.

Nevertheless, even a relatively weak poloidal magnetic field present
in the proto-neutron star may indirectly affect the $T/|W|$ instability.
In the core-collapse supernova scenario, differential rotation naturally
arises inside the stellar core during the collapse (e.g.,
Akiyama \& Wheeler 2005; Ott et al.~2006). This differential rotation
can generate significant toroidal magnetic fields by winding the initial
poloidal field and by magneto-rotational instability
(Akiyama et al.~2003; Obergaulinger et al.~2009). Consider first the
linear winding of the poloidal field $B_p$. The toroidal field
grows in time as $B_\phi\sim B_p\Delta\Omega t$,
where $\Delta\Omega$ is the difference in the rotation rate across the
proto-neutron star. Thus the ratio of magnetic energy
$W_{\rm B}\sim B_{\phi}^2 R^3/6$ and the rotational energy $T\sim 0.2 MR^2
(\Delta\Omega)^2$ increases as
$W_{\rm B}/T\sim B_p^2 R t^2/M$. The time to reach a given $W_{\rm B}/T\equiv f$
is then $t_{\rm twist}\sim (fM/B_p^2R)^{1/2}=\sqrt{f/3}\,R/v_{Ap}$,
where $v_{Ap}=B_p/\sqrt{4\pi \rho}$ is the \Alfven speed associated with $B_{p}$. On the other hand, the
growth time of the low-$T/|W|$ instability is
$t_{\rm grow}\sim 1/\omega_i \sim (R^3/GM)^{1/2}/\hat{\omega}_i$,
where $\hat{\omega}_i$ is the dimensionless growth rate
in units of the Keplerian frequency
($\hat{\omega}_i$ is approximately the vertical axis of the bottom panel in
Fig.~\ref{fig:omega}).
For the low-$T/|W|$ instability to operate before
being suppressed by the large $B_{\phi}$ (generated by twisting the initial
poloidal field $B_p$), we require $t_{\rm twist} \gtrsim t_{\rm grow}$,
i.e.,
\be
B_p \lesssim \hat{\omega}_i\left(f\,{GM^2\over R^4}\right)^{1/2}
\simeq \left(8\times 10^{13}\,{\rm G}\right)
\left({f\over 0.2}\right)^{1/2}
\left({{\hat\omega}_i\over 10^{-3}}\right)
\left({M\over 1.4M_\odot}\right)
\left({R\over 20\,{\rm km}}\right)^{-2}.
\ee
Since $\hat{\omega}_i$ is of order $10^{-3}$ or larger, and the toroidal
field suppresses the instability when $f=W_B/T\sim 0.2$ (see
Fig.~\ref{fig:omega}), we see that an initial poloidal field strong than
$10^{14}$~G can lead to the suppression of the instability.
In other words, when the initial poloidal field is less than $10^{14}~\rm{G}$,
the toroidal field will not grow fast enough by linear winding
so that the low-$T/|W|$ instability still has a chance to develop.

The effect of magneto-rotational instability (MRI) is harder to
quantify. In the linear regime, MRI operates in modes with vertical
structure (i.e., finite vertical wave number), which are independent
from the $T/|W|$-unstable modes studied in this paper. However, the
nonlinear development of MRI may generate significant magnetic fields
(both poloidal and toroidal) on a short timescale (of order the
rotation period). There have been many MHD simulations of core
collapse supernovae (e.g., Ardeljan et al.~2000, 2005; Kotake et
al.~2004; Yamada \& Sawai 2004; Obergaulinger et al.~2006; Burrows et
al.~2007). Most of these simulations cannot resolve
the MRI unless they employ drastically strong initial fields.  It has
been suggested that when MRI saturates, a large fraction of the
kinetic energy in the differential rotation is converted to the
magnetic energy (Akiyama et al.~2003; Obergaulinger et al.~2009),
i.e., $W_{\rm B}/T$ may approach unity on a dynamical time. Our result in
this paper shows that the $T/|W|$ instability is strongly reduced when
$W_{\rm B}/|T|$ reaches 0.2.  Therefore it would be important to
quantify the saturation field of the MRI in proto-neutron stars.
In addition, the MRI can lead to efficient angular momentum transport
in different region of the star. This may also affect the
$T/|W|$ instability. Clearly, these issues must be
resolved in order to evaluate whether the low-$T/|W|$
instability can develop in astrophysically realistic proto-neutron stars.

\section*{Acknowledgments}

This work has been supported in part by NASA Grants NNX07AG81G and
NNX10AP19G, and NSF grants AST 0707628 and AST-1008245.


\label{lastpage}
\end{document}